\documentclass[aps,twocolumn,showpacs,superscriptaddress]{revtex4}
\usepackage{epsfig}
\usepackage{graphicx}
\usepackage{amsfonts}
\usepackage[figuresright]{rotating} 
\usepackage{amssymb}
\usepackage{amsmath}
\usepackage{psfrag}

\def\avg#1{\langle#1\rangle}

\def\be{\begin{equation}} \def\ee{\end{equation}}
\def\bea{\begin{eqnarray}} \def\eea{\end{eqnarray}}
\def\PRB{Phys. Rev. B}
\def\PRA{Phys. Rev. A}
\def\PRL{Phys. Rev. Lett.}
\def\nn{\nonumber}
\def\pp{\parallel}

\begin{document}

\title{Flat bands and Wigner crystallization 
in the honeycomb optical lattice}
\author{Congjun Wu}
\affiliation{Kavli Institute for Theoretical Physics, University
of California, Santa Barbara, CA 93106}
\affiliation{Department of Physics, University of California, San Diego,
CA 92093}
\author{Doron Bergman}
\affiliation{Department of Physics, University of California, 
Santa Barbara, CA 93106}
\author{Leon Balents}
\affiliation{Department of Physics, University of California, 
Santa Barbara, CA 93106}
\author{S. Das Sarma}
\affiliation{Condensed Matter Theory Center, 
Department of Physics, University of Maryland, College Park, MD 20742} 

\begin{abstract}
  We study the ground states of cold atoms in the tight-binding bands
  built from $p$-orbitals on a two dimensional honeycomb optical
  lattice. The band structure includes two completely flat bands. Exact
  many-body ground states with on-site repulsion can be found at low
  particle densities, for both fermions and bosons. 
  We find crystalline order  at $n=\frac{1}{6}$
  with a $\sqrt{3} \times \sqrt{3}$
  structure breaking a number of discrete lattice symmetries. 
  In fermionic systems, if the repulsion is strong enough,
  we find the bonding strength becomes \emph{dimerized} at
  $n=\frac{1}{2}$. Experimental signatures of crystalline order can be detected
  through the noise correlations in time of flight experiments.
\end{abstract}
\maketitle

Optical lattices have opened up a new venue in which to study strongly
correlated systems, with the possibility of precisely controlled
interactions. For example, the superfluid--Mott insulator transition
\cite{greiner2002} for bosons has been observed. Many
interesting states with novel magnetic and superfluid properties in
optical lattices have also been proposed by using high-spin bosons and
fermions \cite{highspin}. In addition, optical lattices provide the
opportunity for studying  another important aspect of
strongly correlated systems, orbital physics, which
plays an important role in metal-insulator transitions,
superconductivity, and colossal magneto-resistance \cite{tokura2000}.

Cold atom systems with orbital degrees of freedom exhibit new features
which are not usually realized in solid state systems. Recently, the
properties of the bosons in the first excited $p$-orbital bands have
attracted a great deal of theoretical attention
\cite{scarola2005,isacsson2005, liu2006, kuklov2006,wu2006,xu2006},
including predictions of a nematic superfluid state 
\cite{isacsson2005}, antiferromagnetic ordering of orbital angular
momentum (OAM) \cite{liu2006,kuklov2006}, a striped
phase of OAM in the triangular lattice \cite{wu2006}, and a bond
algebraic liquid phase \cite{xu2006}. 
Experiments carried out by Browaeys {\it et
al.} \cite{browaeys2005} and K\"ohl {\it et al.} \cite{kohl2005} have
demonstrated the population of higher orbital bands in both bosons and
fermions. 
Recently, F\"olling {\it et al.} has showed the
existence of both the Mott-insulating and superfluid states of the
$p$-band bosons \cite{folling2006}.

The physics of graphene captured by 
the 2D honeycomb lattice has generated 
tremendous interest as a realization of Dirac fermions \cite{graphene1}. 
In graphene, the active bands near the Fermi
energy are ``$\pi$'' type,  composed  of the $p_z$ orbitals
directly normal to the graphene plane.
The other two $p$-orbitals ($p_{x,y}$) lie in-plane, and exhibit
both orbital degeneracy and spatial anisotropy.
However, they hybridize with the $s$-band
and the resulting $\sigma$-bonding band is completely filled.
In contrast, in optical lattices, $p_{x,y}$ bands are well
separated from the $s$-band with negligible hybridization,
giving rise to the possibility of interesting new physics.
The honeycomb optical lattice has already been realized
quite some time ago, by using three
coplanar laser beams \cite{grynberg1993}.

In this article, we study the $p_{x,y}$-orbital physics in the 2D
honeycomb lattice.
We find the lowest energy band is \emph{completely} flat
over the entire Brillouin zone (BZ). 
When the flat band is
partially filled, the effects of interactions are entirely
non-perturbative. For sufficient low densities
the ground states are ``Wigner'' crystals (we
slightly abuse this nomenclature to apply it generally to both bosons
and fermions). We obtain the {\sl exact} many-body
plaquette Wigner crystal state at filling $\avg{n} =\frac{1}{6}$. For
fermionic systems, we obtain additional crystalline ordered states at
higher commensurate fillings. The
noise correlation in the time of flight image should 
detect the order in these states.

We first discuss the single-particle spectrum. 
The optical potential on each site is approximated by a 3D anisotropic
harmonic well with frequencies $\omega_z \gg
\omega_x=\omega_y=\omega_{xy}$, and thus the $p_z$ orbital is well
separated in energy from the $p_{x,y}$-orbitals. 
Since the
honeycomb lattice is bipartite, we denote by $A$ and $B$ its two
sublattices. We define three unit vectors $\hat e_{1,2}= \pm
\frac{\sqrt 3}{2} \hat e_x + \frac{1}{2} \hat e_y$ , $\hat e_3 = -
\hat e_y$ pointing from each $A$ site to its three $B$ neighbors. 
The projection of the $p$-orbitals along the three
$\hat e_{i=1,2,3}$ directions are defined as $ p_i \equiv (p_x {\hat
  e}_x + p_y {\hat e}_y) \cdot {\hat e}_i $.
Only two of them are linearly independent. The kinetic energy  reads 
\bea
H_{0}&=&t_\pp\sum_{\vec r \in A, i=1\sim 3} 
\{ p^\dagger_{\vec r,i} p^{\vphantom\dagger}_{\vec r + a \hat e_i,i}
+h.c. \} -\mu \sum_{\vec r \in A
  \oplus B} n_{\vec r}, \ \ \, \ \ \,
\label{eq:ham0}
\eea 
where the $\sigma$-bonding term $t_\pp$ describes the hopping between 
$p$-orbitals on neighboring sites parallel to the bond direction,
$a$ is the nearest neighbor distance, and $n_{\vec r}=n_
{\vec r,x}+n_{\vec r,y}$ is the total particle number 
in both $p_x$ and $p_y$ orbitals
at the site $\vec r$.
$t_\pp$ is positive due to the odd parity of the $p$-orbitals
and is set to 1 below.
Eq. \ref{eq:ham0} neglects the $\pi$-bonding $t_\perp$ 
which describes the hopping between $p$-orbitals perpendicular
to the bond direction.
Typically $t_\pp/t_\perp\gg 1$ due to the high spatial
anisotropy of the $p$-orbitals.
We have numerically confirmed this for the optical potential  
realized in experiment
$V(\vec r)=V_0\sum_{i=1\sim 3} \cos (\vec p_i \cdot \vec r)$ where
$\vec p_i =p_0 \hat e_i$ and $p_0=\frac{4\pi}{3 a}$ \cite{grynberg1993}.
Namely, $t_\perp \approx 0.02 t_\pp$ and $t_\pp\approx 0.24 E_R$ 
at $V_0/E_R=5$ ($E_R$ is the recoil energy).

\begin{figure}
\centering\epsfig{file=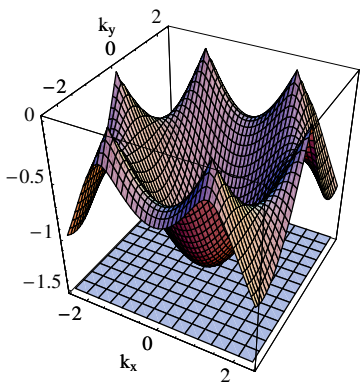,clip=1,width=0.55\linewidth,angle=0}
\caption{Dispersion of the two-lowest $p_{x,y}$-orbital bands
  $E_{1,2}$. 
  The band $E_1$ is completely flat, while $E_2$ exhibits Dirac points at
  $K_{1,2}=(\pm\frac{4\pi}{3\sqrt 3 a},0)$. The other two bands are
  symmetric with respect to $E=0$.}
\label{fig:hnycmb_band}
\end{figure}

The band structure  contains both flat bands and Dirac cones.
Each unit cell consists of two sites, each of which contains two 
orbitals $p_{x,y}$, resulting in four bands. 
The BZ takes the shape of a regular hexagon with edge length
$4\pi/(3\sqrt 3 a)$.
The dispersion of all the  bands is symmetric with respect to 
the zero energy because of the bipartite nature of the lattice.
The band structure consists of
$
E_{1,4}(\vec k)= \mp \frac{3}{2} ,
E_{2,3} (\vec k)= \mp \frac{1}{2} \sqrt{3+2\sum_i \cos \vec k\cdot \vec b_i}
$
where $\vec b_{1}= a \left( \hat e_2 -\hat e_3 \right)$, 
$\vec b_2= a \left( \hat e_3-\hat e_1 \right)$
and
$\vec b_3= a \left( \hat e_1-\hat e_2 \right)$.
Wie show only the two lowest bands
$E_{1,2}$ in Fig. \ref{fig:hnycmb_band}, which touch at
the Brillouin zone center.
The bottom and top bands turn out to be completely flat.
The two middle bands are dispersive with a width determined by
$t_\pp$, which are the same as in graphene
with two non-equivalent Dirac points located at 
$\vec K_{1,2}=(\pm \frac{4\pi}{3\sqrt 3 a},0)$.
We will not repeat the extensively studied Dirac cone physics here,
but rather focus on the new features brought by the flat bands instead.

\begin{figure}
\psfrag{R}{$\vec R$}
\centering\epsfig{file=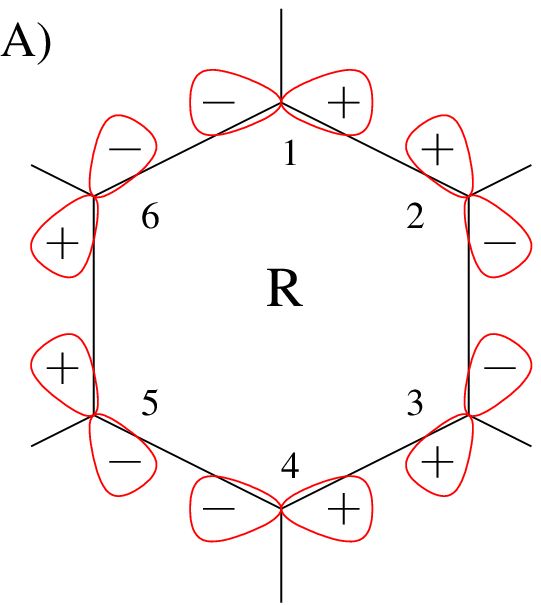,clip=1,width=0.4\linewidth,angle=0}
\psfrag{R1}{$\vec R_1$}
\psfrag{R2}{$\vec R_2$}
\psfrag{R3}{$\vec R_3$}
\psfrag{R}{$\vec R$}
\hspace{3mm}
\centering\epsfig{file=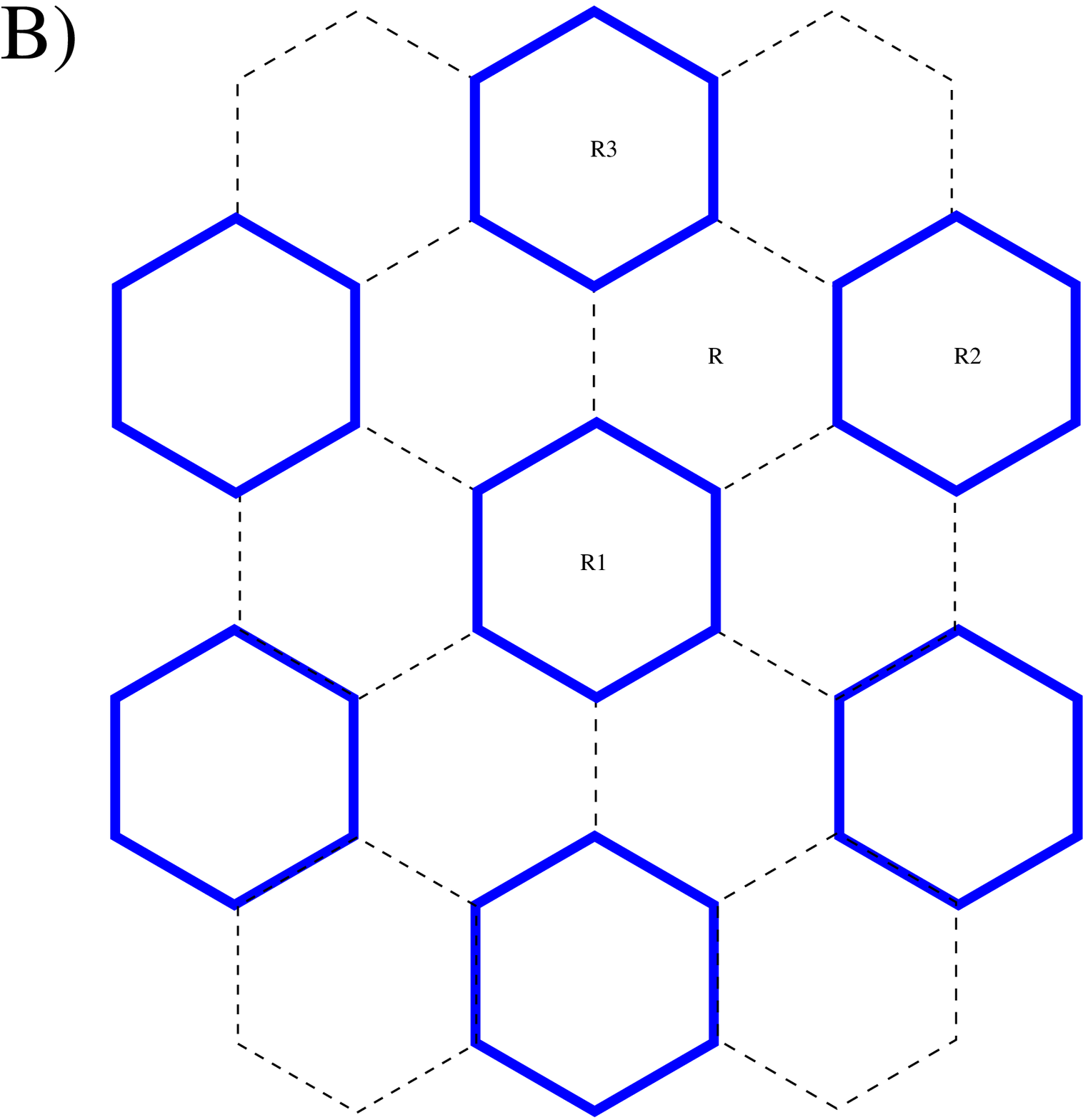,clip=1,width=0.45\linewidth,angle=0}
\caption{
A) The Wannier-like localized eigenstate for the lowest band. 
The orbital configuration at each site is oriented along 
a direction tangential to the closed loop on which the particle is delocalized.
B) The configuration of the close-packed Wigner-crystal
state for both bosons and fermions at $n=1/6$. 
Each thickened plaquette has the same
configuration as in A).
}
\label{fig:closepack}
\end{figure}

The degeneracy of all momentum eigenstates in the flat bands
allows us to take any superposition of these states, and in particular,
to construct eigenstates that are \emph{localized}.
As depicted in Fig. \ref{fig:closepack} A, for each hexagon plaquette 
$\vec R$, there exists one such eigenstate for the bottom band 
\bea
|\psi_{\vec R}\rangle&=&\sum_{j=1}^6 (-)^{j-1}\Big\{ \cos\theta_j
|p_{j,x }\rangle 
-\sin \theta_j |p_{j,y}\rangle \Big\} \label{eq:wannier}
\eea
where $j$ is the site index and $\theta_j=(j-1)\frac{\pi}{3}$. 
The orbital configuration on each site is perpendicular to the 
links external to the hexagonal loop.
This orbital orientation,
together with destructive interference of hopping amplitudes, 
prevent the particle from ``leaking'' off of the plaquette.
The states $|\psi_{\vec R}\rangle$ are all
linearly-independent apart from
one constraint $\sum_{\vec R} |\psi_{\vec R}\rangle=0$ under
periodic boundary conditions.
The Bloch wave states are constructed as
$|\psi_k\rangle_1=\frac{1}{\sqrt {N_k}} \sum_k e^{i\vec k \cdot \vec R}
|\psi_{\vec R}\rangle (\vec k \neq (0,0))$, with the normalization factor
$N_k=\frac{8}{3} (3-\sum_i \cos \vec k \cdot \vec b_i)$.
The doubly degenerate eigenstate at $\vec k =(0,0)$
can not be constructed from the above plaquette states.
They are $|\psi_{\vec k=(0,0)}\rangle_{1,2}=
\sum_{\vec r\in A} |p_{x(y), \vec r} \rangle-
\sum_{\vec r\in B} |p_{x(y), \vec r} \rangle$.

Because of the orbital degeneracy, the onsite interaction for 
spinless fermions remains Hubbard-like as
\bea
H_{int}= U \sum_{\vec r } n_{\vec r, x} n_{\vec r, y}.
\label{eq:hamint}
\eea 
In order to enhance $U$ between neutral atoms, we may use the
$^{53}$Cr atom which has a large magnetic moment of $6\mu_B$ (Bohr
magneton), and polarize the spin with an external magnetic field. 
The length scale of $p_{xy}$-orbitals $l_{x,y}
=\sqrt{\hbar/m\omega_{x,y}}$ is typically one order smaller than $a$. 
For example, we estimate that $l_{x,y}/a\approx 0.2$ at $V_0/E_R=5$.
Assuming strong confinement in the $z$-axis $l_z\ll l_{x,y}$, 
the vector linking two atoms in $p_x$ and $p_y$ orbits almost 
lies in the plane.
$U$ can be adjusted from repulsive to attractive 
by tuning the polarization direction from perpendicular to parallel
to the $xy$-plane.
We will only discuss repulsive $U$ below. 
We estimate that $U$ can easily reach the order of $E_R$ 
which is much larger than $t_\pp$. 
The off-site interactions are small and decay
with distance as $1/r^3$, and thus are neglected.
For example, the nearest neighbour interaction is at the order of 
$(\frac{l_{x,y}}{a})^3 U\approx 10^{-2} U$.
The on-site interaction for $p$-band bosons is given 
in Ref. \cite{liu2006,wu2006}.

When the flat band is partially filled the effect of interactions 
is non-perturbative. 
At sufficiently low particle density $n \leq 1/6$, both the interactions
and the kinetic energy can be minimized simultaneously.
This can be realized if the individual particles localize into the 
plaquette states depicted in Fig. \ref{fig:closepack} A. These are exact 
eigenstates of the flat band, and therefore minimize kinetic energy.
Any arrangement of the plaquette states where they do not overlap
will cost no interaction energy.
These localized plaquette states therefore constitute \emph{exact}
many-body ground states.
The maximum density (close packed) arrangement of non-overlapping plaquettes 
has filling $n=1/6$, and has the structure depicted in
Fig. \ref{fig:closepack} B. 
The $\sqrt{3}\times \sqrt{3}$ structure, breaks the lattice translation
symmetry and is three-fold degenerate. 
At zero temperature, the particle density {\sl
 jumps} from $0$ to $1/6$ as the chemical potential is increased
through $\mu=-3/2$. 
Similar flat band behavior has been studied in the magnon spectra of a
number of frustrated magnets in
a large magnetic field \cite{zhitomirsky2004} near full
polarization of the magnet. The degenerate states for the Kagome
antiferromagnet are in exact one to one correspondence to those of our
model. Thus the two models have identical thermodynamics (if
fluctuations are restricted to these states, as appropriate for
$k_{\scriptscriptstyle B}T \ll t_\parallel, U$). This is described by
the classical hard hexagon model\cite{zhitomirsky2004}, which exhibits a
second order thermal phase transition in the 3-state Potts model
universality class, breaking translational symmetry, when the fugacity
of the hexagon $z_c=(11+\sqrt 5)/2$. 
In this state, the atoms do not touch each other,  
thus particle statistics do not play any role. The Wigner
crystal is also expected to appear for bosons in this optical lattice at 
filling $n=\frac{1}{6}$.
In contrast, Wigner crystal is not possible in graphene systems
\cite{dahal2006}.

\begin{figure}
\centering\epsfig{file=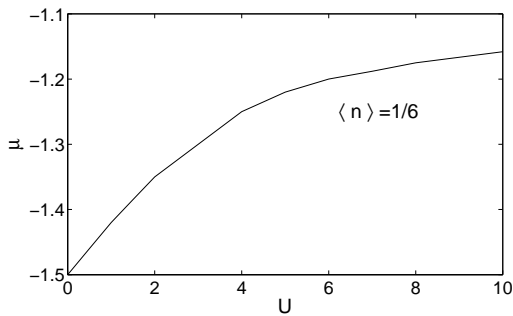,clip=1,width=0.8\linewidth,angle=0}
\caption{The phase boundary of the incompressible plaquette Wigner 
crystal state of spinless fermions at $\avg{n}=\frac{1}{6}$. 
}\label{fig:gap}
\end{figure}

\begin{figure}
\centering\epsfig{file=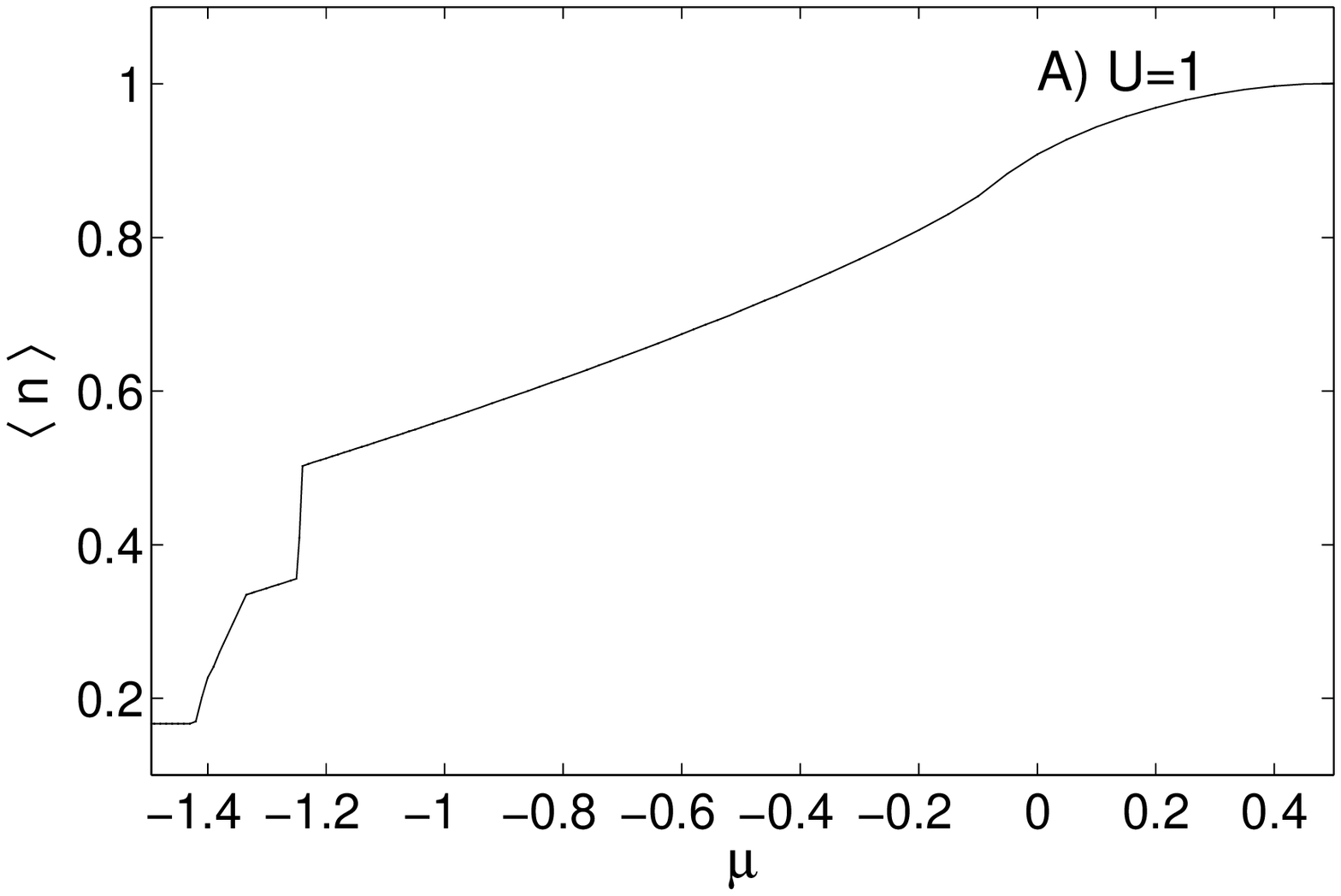,clip=1,width=0.70\linewidth,angle=0}
\centering\epsfig{file=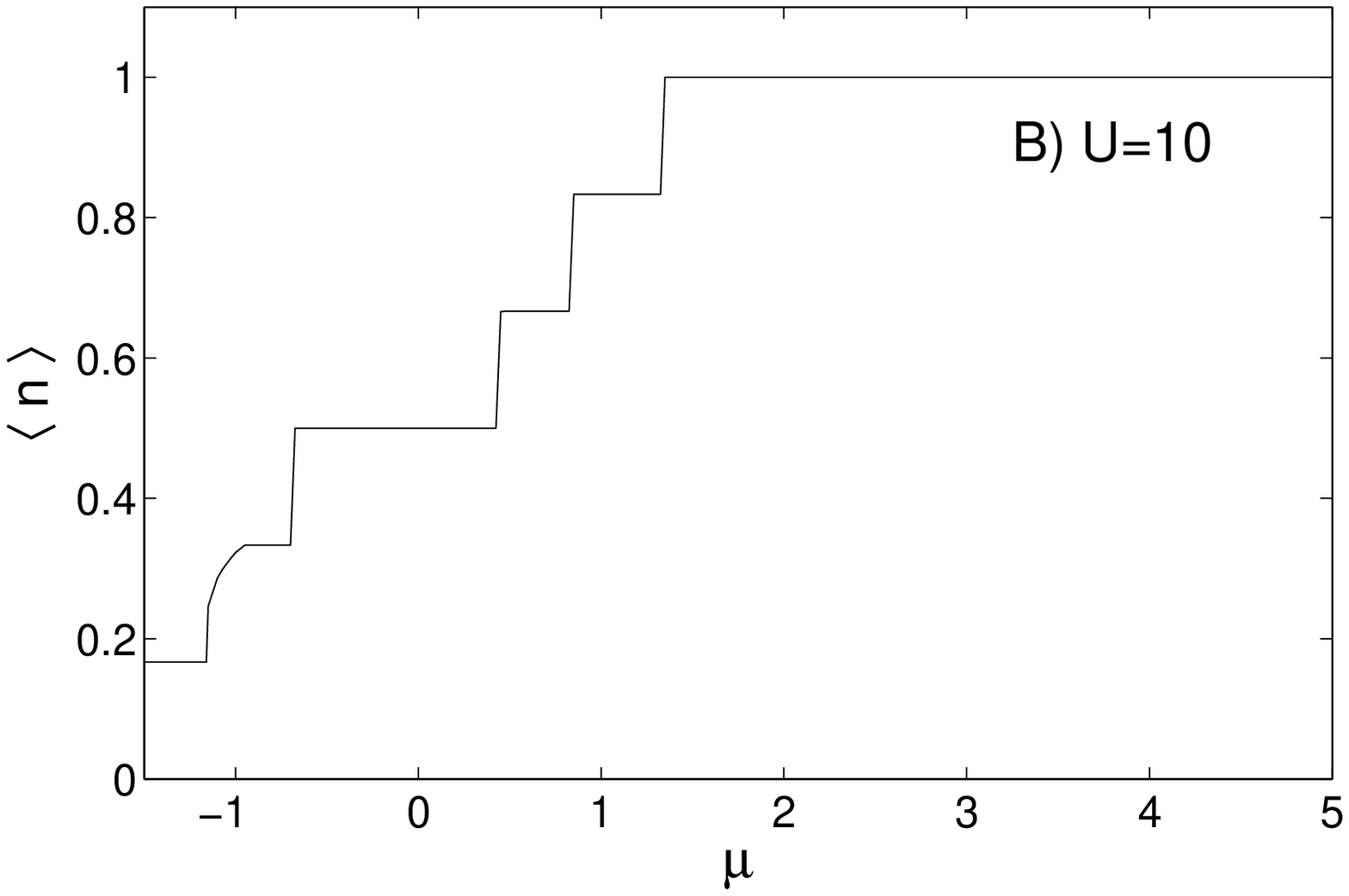,clip=1,width=0.70\linewidth,angle=0}
\caption{The filling $\avg{n}$ vs. the chemical potential $\mu$ for
  spinless fermions for weak A) and strong B) interactions. Due to
  particle-hole symmetry, only the part with $\mu$ from the band bottom
  $-\frac{3}{2}t_\pp$ to $U/2$ is shown. Only one plateau appears in A)
  at $n=\frac{1}{6}$, while a series of plateaus appear in B) at $n=1/6,
  1/3, 1/2, 2/3, 5/6, 1$. }\label{fig:filling}
\end{figure}

When the filling $n>1/6$, exact solutions are no longer available
for the interacting Hamiltonian. For simplicity, 
in this regime we shall only discuss the case of spinless fermions,
within a mean field 
treatment. 
We decouple Eq. \ref{eq:hamint}
in both the direct and exchange channels as
\bea
H_{mf,int}&=&\frac{U}{2}\sum_{\vec r \in A \oplus B}
\Big\{ n_{\vec r, x} \Big[\avg{n_{\vec r}} -\avg {n_{\vec r,3}} \Big]
+n_{\vec r, y}\Big[\avg{n_{\vec r}} \nn \\
&&\hspace{-10mm} +\avg {n_{\vec r,3}}\Big]
-p^\dagger_{\vec r x} p^{\vphantom\dagger}_{\vec r y} \Big[\avg {n_{\vec r,1}}
-i\avg {n_{\vec r,2}}\Big]
-h.c. \Big\},
\label{eq:mft}
\eea
and solve it self-consistently.
Here $n_{1,2,3}$ are the pseudo-spin operators defined as
$
n_{\vec r, 3} = \frac{1}{2} (p^\dagger_{\vec r x}p_{\vec r x}-
p^\dagger_{\vec r y}p_{\vec r y}), 
n_{\vec r, 1} =\frac{1}{2} (p^\dagger_{\vec r x} p_{\vec r y}+h.c.),
n_{\vec r, 2} =\frac{1}{2i} (p^\dagger_{\vec r x} p_{\vec r y}-h.c.).
$
The $n_{\vec{r},1},n_{\vec{r},3}$ operators are time-reversal invariant,
and describe the preferential occupation of a ``dumbbell-shaped'' 
real $p$-orbital orientation; $n_{\vec{r},2}$ is the
orbital angular momentum, and is time-reversal odd.
At the mean field level $\langle n_{\vec{r},2}\rangle$ is zero. 
In order to obtain the plaquette order in Fig. \ref{fig:closepack}, we take
the six sites around a plaquette as one enlarged unit cell in the mean
field calculation.

The numerically determined phase boundary between the $1/6$ state and
a compressible phase at higher density is shown in Fig. \ref{fig:gap}.
The charge gap grows roughly linearly with
$U$ in the weak interaction regime, and saturates at 
a value comparable to $t_\pp$ in the strong interaction regime.
This can be understood as follows: 
in the weak interaction case, we choose a plaquette $\vec R$ 
which is adjacent to three occupied plaquettes $\vec R_{1,2,3}$
as depicted in Fig. \ref{fig:closepack}. B,
and put an extra particle in it.
The cost of the repulsion is $\frac{U}{6}$.
In the strong coupling case, we put the particle
into an excited state of the occupied plaquette $\vec R_1$
while fixing the orbital configuration on each site. 
Because fermions are spinless, the cost of energy comes from
the kinetic part with the value of $\frac{3}{4}t_\pp$.
Thus the charge gap $\Delta <\mbox{min}(\frac{1}{6} U, \frac{3}{4}t_\pp)$
which agrees with the numeric result.

The curves of the filling $n$ vs. $\mu$ in both 
the weak and strong coupling regimes
are depicted in Fig. \ref{fig:filling}.
In both cases, the plaquette Wigner state appears as a plateau
with $\avg{n}=\frac{1}{6}$.
In the weak coupling regime ($U/t_\pp=1$), as $\mu$ passes the charge gap,
$\avg{n}$ increases quickly corresponding to filling up
other states in the flat band.
Due to the background crystal ordering, those states
are no longer degenerate, but develop weak dispersion.
A significant reduction in density of states (DOS) occurs 
at the approximately commensurate filling of $\avg{n}\approx \frac{1}{3}$, 
but it is not a strict plateau.
At $\avg{n}>1/2$, after the flat band is completely filled, the
interaction effects are no longer important. 
Near half-filling, the DOS vanishes where the physics is dominated
by the Dirac cones.

\begin{figure}
\centering\epsfig{file=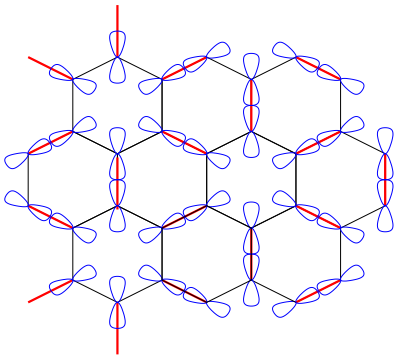,clip=1,width=0.6\linewidth,angle=0}
\caption{The dimerized state of spinless fermions with filling
  $\avg{n}=\frac{1}{2}$. Each thickened (red) bond corresponds a dimer
  containing one particle. }\label{fig:dimer}
\end{figure}

The physics changes dramatically in the strong coupling regime. 
We looked specifically at ($U/t=10$), where we found a series 
of plateaus appear at commensurate
fillings $\avg{n}=\frac{i}{6} (i=1\sim 6)$.
The large charge gap at $\avg{n}=1$ is of the order of $U$.
The other plateaus with $\avg{n}<1$ are the novel features
the $p$-orbital model presents.
In addition to the plaquette bond order at $\avg{n}=\frac{1}{6}$ 
discussed above, rich structures including dimer and trimer 
bond orders appear at other plateaus with $\avg{n}<1$.
We will present these patterns in detail in a future publication.
Here a dimer refers to a superposition of 
the two states of two sites
where one is occupied, while the other is empty.
The dimerized state at $\avg{n}=\frac{1}{2}$ 
is illustrated in Fig. \ref{fig:dimer}.
Each dimer is represented as a thickened bond where
the orbital configuration is along the bond.
We check that the bonding energy for the thickened bond is 
approximately $0.95t_\pp$ at $U/t_\pp=10$ which is about one
order larger than that of other bonds.

Next we discuss the effect of adding a small $\pi$-hopping $t_\perp$ term. 
Then the lowest energy band acquires weak dispersion, with a small
band width of $t_\perp$. This associated kinetic energy cost for 
the $n=\frac{1}{6}$ plaquette state is of order $t_\perp$ per particle.
A stability condition of this state can therefore be roughly estimated as
$\frac{U}{6}>t_\perp$. We have checked numerically that the
$\frac{1}{6}$-plateau survives at $U>t_\pp$ with the value of
$t_\perp=0.1 t_\pp$. Other plateaus appearing in the strong coupling
regime are not sensitive to small $t_\perp$.

The plaquette Wigner crystal phase in Fig . \ref{fig:closepack} B
should manifest itself in the
noise correlation for the time of flight (TOF) signals.
In the presence of the plaquette order, the reciprocal lattice
vector for the enlarged unit cell becomes
$\vec G_1=(\frac{4\pi}{3\sqrt 3 a},0)$ 
and $\vec G_2=(\frac{-2\pi}{3\sqrt 3 a}, \frac{2\pi}{3a})$.
The correlation function is defined as
$C_t(\vec r, \vec r^\prime)=\avg{n(\vec r) n(\vec r^\prime)}_t 
-\avg{n(\vec r)}_t \avg{n(\vec r^\prime)}_t.
$
After a spatial averaging and normalization, we find
\bea
C_t(\vec d)=\int d \vec r \frac{C_t(\vec r +\frac{\vec d}{2},
\vec r -\frac{\vec d}{2})}
{\avg{n(\vec r+ \frac{\vec d}{2})}_t \avg{n(\vec r
-\frac{\vec d}{2})}_t}\propto \pm\sum_{\vec G}\delta (\vec k -\vec G)
\eea
where $'+'$ ($'-'$) is for bosons (fermions) respectively,
 $\vec k=m \vec d/(\hbar t)$, and $\vec G= m \vec G_1 
+ n \vec G_2$ with $m,n$ integers.

There are numerous directions open for further exploration.
Some interesting variations on the model are to consider spin-{\sl ful}
fermions, for which there is a possibility of ferromagnetism as a result of the flat bands,
or attractive interactions, in which pairing and the BCS-BEC crossover
in the flat band might prove interesting. Most intriguing is the
possibility of exotic incompressible states analogous to the Laughlin
liquid in FQHE. These cannot be captured
within the mean-field approximation used here for $n>1/6$. If one could
devise appropriate variational liquid states projected into the flat
band, these could be compared energetically with the Wigner crystals
found here. Given the richness and surprises encountered in the FQHE,
flat band physics in optical lattices appears rife with possibility.
The comparison with the fermion condensation where
the flat band arises due to strong interactions will also be
interesting \cite{khodel}.

C. W. thanks helpful discussions with A. Carlsson, L. M. Duan, and
Z. Nussinov.
C. W. is supported by the NSF No. Phy99-07949. D.~B. and L.~B. are
supported by NSF-DMR-04-57440 and the Packard foundation. S. D. S. is
supported by LPS-NSA and ARO-DTO.


\end{document}